# STABILIZING SLIDING MODE CONTROL DESIGN AND APPLICATION FOR A DC MOTOR: SPEED CONTROL

Ahmed Rhif

Department of Electronic, High Institute of Applied Sciences and Technologies, Sousse, Tunisia

(Institut Supérieur des Sciences Appliquées et de Technologie de Sousse, Tunisie)

E-mail: ahmed.rhif@gmail.com

## ABSTRACT

*The regulation by sliding mode control (SMC) is recognized for its qualities of robustness and dynamic response. This article will briefly talk about the regulation principles by sliding mode as well as the application of this approach to the adjustment of a speed control DC motor bench using the TY36A/EV unit. This unit, from Electronica Veneta products, uses a PID controller to control the speed and position of the DC motor. Our purpose is to improve the set time answer and the robustness of the system when disturbances take place. The experimental results show very good performances of the proposed approach relatively to the PID.*

## Keywords
*Sliding mode control, Lyapunov stability, PID controller, DC motor.*

## 1. INTRODUCTION

The electrical servomechanisms for DC motors [1] are, nowadays, widely used in several robotic applications and electrical engineering. Concern the position controls, they are mostly employed in machine tools where we require a high precision between the tool and the piece that would be machined. The device used in this work is carried out within the laboratory "Control and Regulation" of the Department of Electronic Engineering of the High Institute of Applied Sciences and Technology of Sousse (Institut Supérieur des Sciences Appliquées et de Technologie de Sousse I.S.S.A.Tso). This device is composed by two elements: the module G36A/EV and the mechanical process TY36A/EV.
The board mod.G36A/EV has been designed and carried out for the training of technicians with high level of knowledge on DC motor controls. The board mod.G36A/EV, carried out with industrial components: circuits and techniques, exist together with the process unit TY36A/EV for the development and the training of experimental programs enables the development of exercises concerning the:

- Study of electrical servomechanisms for DC motors;
- The speed and position control of a DC motor.
- Study of DC motors;

Algorithms of the traditional control using the regulators with Proportional, Integral and Derivative actions allow the control of non disturbed linear processes and with constant parameters [2, 3, 4].  When the process controlled is subjected to disturbances and parameters variations, an adaptive solution would be needed, which by readjustment of the regulators parameters, that makes possible to preserve performances fixed in advance in the presence of disturbances and of parameters variations. This solution presents the disadvantage of the complex implementation.

Thus, it is possible to use another simple solution, using particular class of control devices, called "Variable System Structure (VSS)". These systems were the important work subject for a long time in Japan with F. Harachima [5, 6], in the United States with Slotine and with the ex-Soviet V. Ukin [5] and this starting from theoretical work of the Soviet mathematician A.F. Filipov. The recent interest granted to this control technique is due to availability of the powerful electronic components and to the very developed microprocessors [7].

The paper is structured in the following way: the first part is devoted to the system description then to the development of the state model of the process based on the DC motor.  Then, we define the sliding mode approach as well as the theorical synthesis, the electronic design and the realization of the sliding surface and of the process control law. The experimental results are presented in the section 5.

## 2.  THE SYSTEM DESCRIPTION

The process TY36A/EV based on a DC Motor is controlled by the electronic device mod.G36A/EV from Electronica Veneta products (Figure 1). The main circuit blocks of the board mod.G36A/EV are:

- • Set-point;
- • Error Amplifier;
- • PID controller;
- • Signal conditioners for transducers.

The Set point (speed and position) fixing is made via an inner voltage reference variable by a rotary potentiometer. The different PID controller actions are independently calibrated using three rotary potentiometers. The controlled speed of the DC motor, in a closed loop transduced by the optoelectronic device, is displayed in a seven segment and four digit displays. The connection between this module and the process unit TY36A/EV is made via two terminals, which provide power, and an eight pole DIN socket that is used to connect the incoming signals from the speed and position transducers. Finally, a data acquisition software and hardware are used for supervision from Personal Computer.

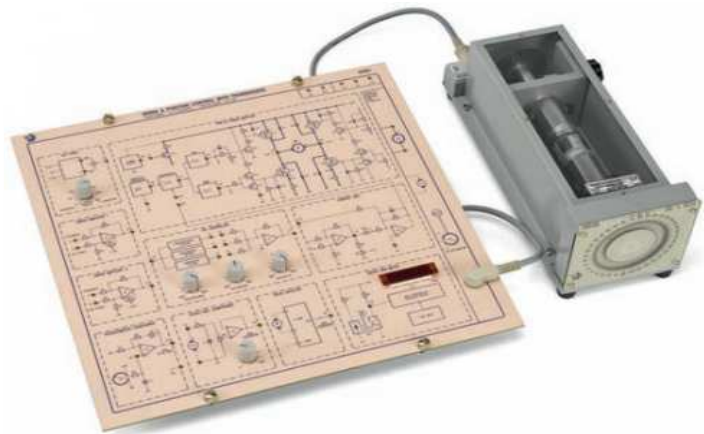

Figure 1 The electromechanical process

*The process unit TY36A/EV*
The mechanical process TY36A/EV is composed by:

- DC permanent magnets motor;
- Devices for load variation (disturbances);
- Potentiometer position transducer;
- Tachometric and optoelectronic speed transducers.

This process, with the rotation of the DC permanent magnets motor, enables the generation of speed and position outputs. The speed signals reach, via the eight poles DIN cable, the board mod.G36A/EV where they are processed. Other side, the angular position of the rotor can be directly read on the external position indicator ranging between 0° and 360°.

## 3. PROCESS MODELLING

The system based on the DC motor is considered as a benchmark for the study and analysis of speed and position problems (Figure 2). This device, by "ElectronicaVeneta", allows us to examine the stability, precision, speed request and robustness of the actuator controlled by a PID [8, 9].
This mechanical process TY36A/EV is composed of:

- DC permanent magnet motor;
- Optoelectronic transmission sensor;
- Transparent and opaque disk for fork optoelectronic transducer;
- Tachogenerator fixed to the motor axis;
- Potentiometric transducer;
- Epicycloidal planetary gearing;
- Angular position indicator for the angular position.

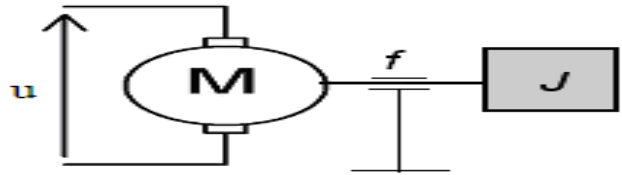

Figure 2 The TY36A/EV unit conception

In this way, the process is represented by the following mathematical system (1).

$$\begin{cases} u(t) = e(t) + Ri(t) + L\dfrac{di(t)}{dt} \\[2mm] c_m(t) - c_r(t) = f\omega(t) + j\dfrac{d\omega(t)}{dt} \\[2mm] c_m(t) = k_1 i(t) \\[2mm] e(t) = k_2 \omega(t) \end{cases} \qquad (1)$$

with $k_1$ and $k_2$ are the conversion constants (torque coefficients), where $u$ is the motor armature voltage, $i$ is the armature current, $e$ is the back electromotive-force voltage, $\omega$ is the shaft speed, $c_m$ the mechanical couple and $c_r$ the resistant torque [10].

To determine the system transfer function, we applied Laplace transformation which gives (2).

$$\begin{cases} U(t) = E(s) + RI(s) + LsI(s) \\ C_m(s) - C_r(s) = F\Omega(s) + Js\Omega(s) \\ C_m(s) = K_1 I(s) \\ E(s) = K_2 \Omega(s) \end{cases} \qquad (2)$$

Table 1 presents the DC permanent magnet motor characteristics.

| Rotation speed | 4000 RPM |
|---|---|
| Nominal voltage | 24 V |
| Armature resistance R | 5.5 Ohm |
| Armature inductance L | 2.8 mH |
| The inertia moment J | 0.0163 Kg.m2.s2 |
| The friction forces f | 0.2 N.m.s |
| Weight | 5 kg |

Table 1 DC motor parameters

These equations could be represented in the diagram block shown in the figure 3.

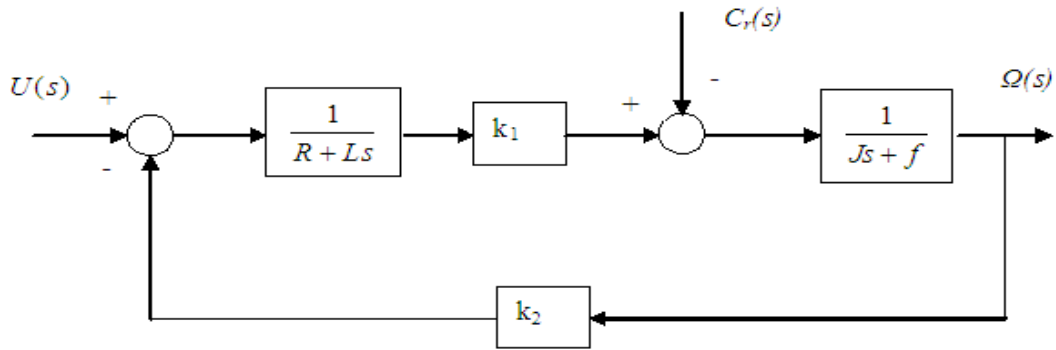

Figure 3 The Bloc diagram of a DC motor regulation

From equation (2), we are able to conclude the transfer function (3) then the state model (5) of the system TY36A/EV.
In this example, we consider that $k_1 = k_2 = k$.

$$\Omega(s) = \frac{k}{Rf + k^2 + (RJ + Lf)s + LJs^2} U(s) - \frac{R + Ls}{Rf + k^2 + (RJ + Lf)s + LJs^2} C_r(s) \qquad (3)$$

Consider the state system (4).

$$\begin{cases} \dot{X} = AX + Bu \\ Y = CX + Du \end{cases} \qquad (4)$$

In this work we consider that $f$ and $c_r$ are null, from where we conclude (5).

$$\begin{bmatrix} \dfrac{di}{dt} \\ \dfrac{d\omega}{dt} \end{bmatrix} = \begin{bmatrix} -\dfrac{R}{L} & \dfrac{k}{L} \\ \dfrac{k}{J} & 0 \end{bmatrix} \begin{bmatrix} i \\ \omega \end{bmatrix} + \begin{bmatrix} \dfrac{1}{L} \\ 0 \end{bmatrix} u$$

$$\omega = \begin{bmatrix} 0 & 1 \end{bmatrix} \begin{bmatrix} i \\ \omega \end{bmatrix} + \begin{bmatrix} 0 \end{bmatrix} u \qquad (5)$$

with

$$A = \begin{bmatrix} -\dfrac{5.5}{0.0028} & -\dfrac{1}{0.0028} \\ \dfrac{1}{0.0163} & 0 \end{bmatrix}, \qquad B = \begin{bmatrix} \dfrac{1}{0.0028} \\ 0 \end{bmatrix} , \quad C = [0\ 1] \text{ and } D = [0]$$

## 4. THE SLIDING MODE CONTROL

### 4.1 The sliding mode approach

In this kind of regulation (SMC), the system state defines the position control unit. The idea is to divide the state space by a sliding surface, which delimit two spaces corresponding in two possible states control unit. The system state will converge to the sliding surface in a first time, and then it will slide on this surface to reach the desired state [11, 12, 13, 14, 15, 16].

Stabilization on the sliding surface is obtained by using a commutation between two extreme values. This principle of commutation law consists on the use of a discontinuous control having as function the state maintain on the sliding surface and the disturbances rejection.

On the surface, the system dynamics is independent of the initial process. These concepts of local stability will be shown by taking account of the stability principle according to the Lyapunov criterion [17, 18, 19].

### 4.2 A sliding mode control synthesis

In this part, we will present the necessary mathematical equations and details for the synthesis of a sliding mode control [20]. For this type of study, we consider the case of commutation on the control unit with addition of the equivalent control (Figure 4). Other ways, we present the electronic design of this sliding surface that may be implemented to control the process in the testing phase (Figure 5).

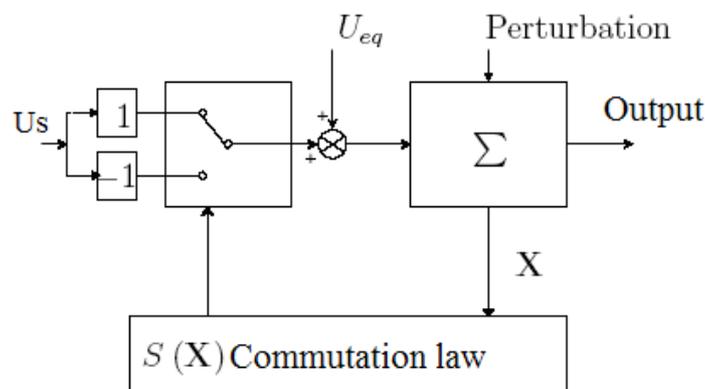

Figure 4 Control unit with addition of the equivalent control

Consider $s$ a sliding surface such as:

$$s = C\hat{x} \tag{6}$$

with $\hat{x} = x_d - x_r$

To carry out the system on the sliding surface $s$, we have to select a discontinuous control which commutates between two extremes values: $u_s = -ksign(s)$ , with $k>0$. When the system reaches the surface, the process control $u$ is equal to the equivalent control $u_{eq}$ (7).

$\dot{s} = C\dot{x} = CAx + CBu_{eq}$

$$\Leftrightarrow u_{eq} = -(CB)^{-1}CAx \tag{7}$$

we conclude that the global control of the system considering the two phases (reaching the sliding surface and the sliding phase to the equilibrium state) is represented in (8).

$$u = -(CB)^{-1}CAx - ksign(s) \tag{8}$$

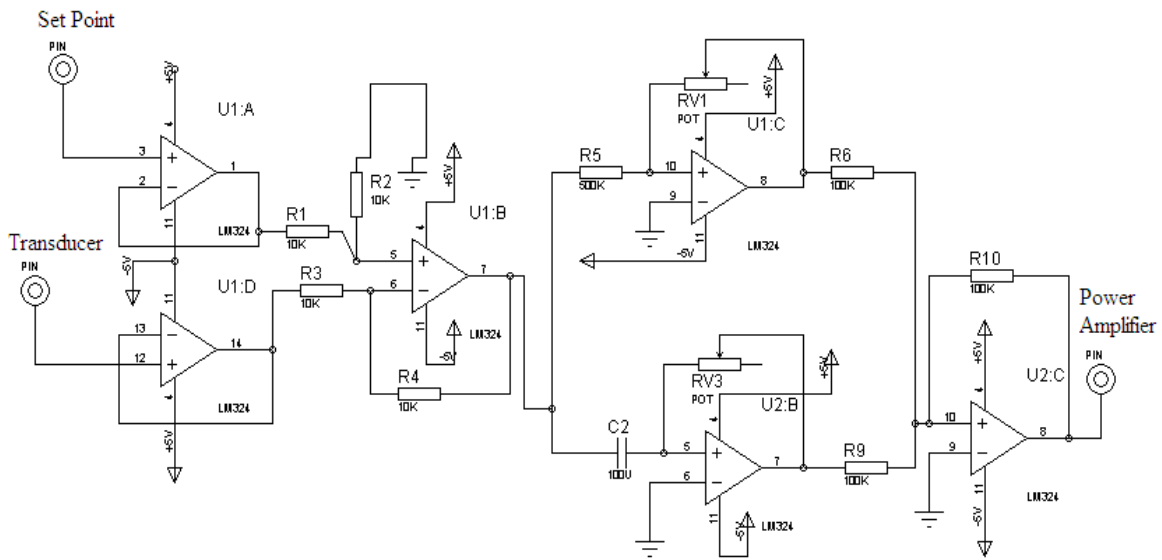

Figure 5 The electronic conception of the sliding surface

To ensure the system stability carried out by this control, we consider the Lyapunov candidate function $v = \frac{1}{2}s^2$, $\dot{v} = s\dot{s}$, then we have to prove that $\dot{v} < 0$.

$\dot{s} = C\dot{x} = CAx + CBu = -CBksign(s)$

to ensure $\dot{v} < 0$ , we must have $CBk>0$.

The implementation structure of the sliding mode control previously synthesized for the speed control is represented in Figure 6.

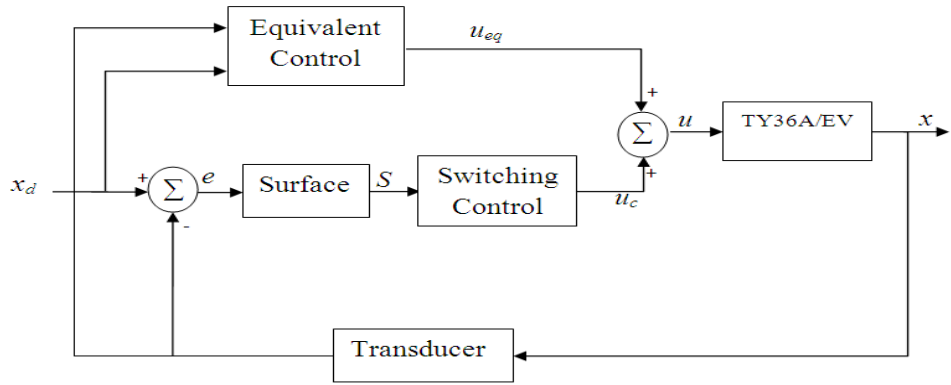

Figure 6 Sliding mode control bloc representation

# 5. EXPERIMENTAL RESULTS

The experimental results of this work are shown by figures 7 to 13. In the beginning, we start by fixing the set point (1V≈ 660 RPM). The first test is using the PID controller which exists in the original system. This controller allowed us to reach the desired DC motor speed but with late response (after 18s, Fig. 7). Also, the PID controller showed a non robustness to disturbances (Fig.7-8); when we inject the friction coefficient by using the mechanical handle (Fig.9), the motor speed decrease significantly (80%) then reach again the desired point. We notice also, that the control level (Fig.8) is quietly high (U≈ 15) and it presents a sharp variation due to the disturbance. By using the sliding mode control, the system response is faster (3s) and the output is more precise and stable (Fig.10). The disturbance is not very effective in this case; the DC motor speed decrease only by (1.5%). The control level and commutation frequency are also lower. Other way, the perturbation that exists in the sliding surface is due to the abrupt injection of the friction force *f*. we conclude that the sliding mode control is more robust and present more speed response that gives the desired evolution of the system output.

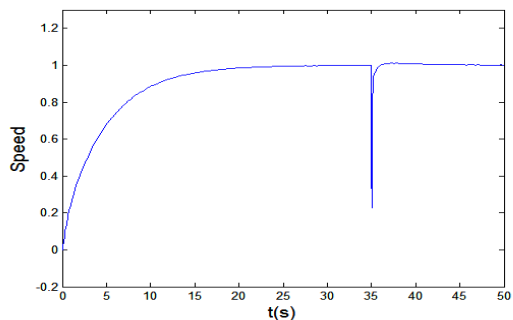

Figure 7 Output evolutions by PID

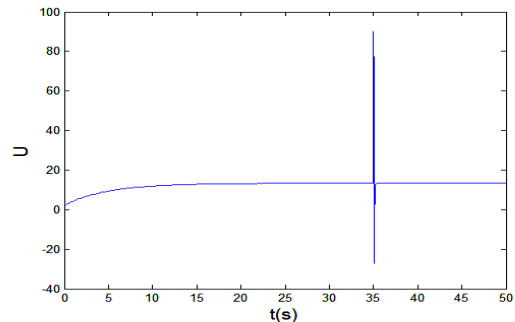

Figure 8 Control evolutions by PID

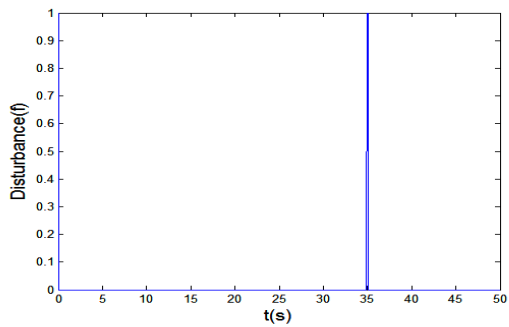

Figure 9 Disturbance Signal

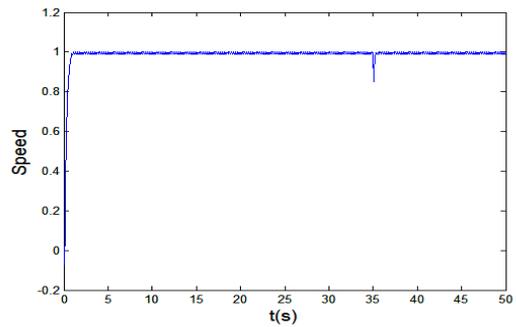

Figure 10 Output evolutions by SMC

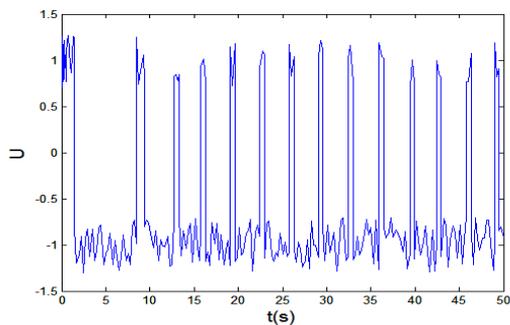

Figure 11 The control evolutions by SM

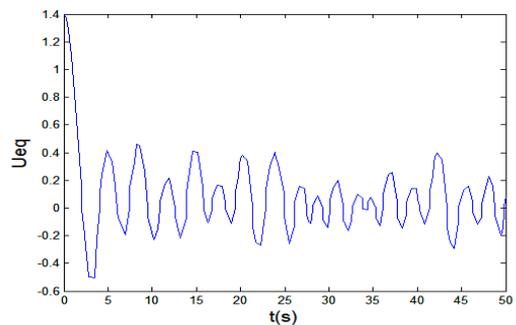

Figure 12 The equivalent control

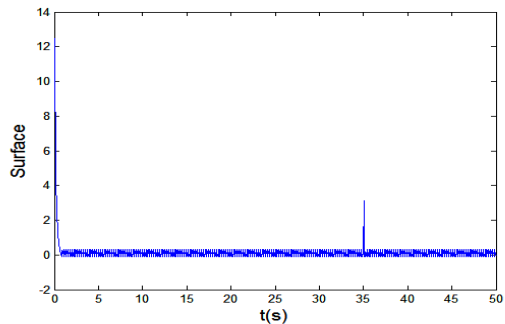

Figure 13 Sliding surface evolutions

## 6. CONCLUSION

In this work, we approached the synthesis method of a stabilizing control law by sliding mode using a nonlinear sliding surface. We start by designing the sliding surface and representing the implementation of the electronic card that may be connected to the TY36A/EV unit. Then, multiple tests were done to control the process and to show differences between the old controller used, PID, and our new proposition that gives better results in lack and in the presence of disturbances. We conclude that the sliding mode control is more robust and gives quick reply to the system.

## 7. ACKNOWLEDGEMENTS

The author wants to thank the High Institute of Applied Sciences and Technologies (ISSATso) in particular the Department of Electronic Engineering for the equipments granted to carry out this project. The author likes also to thank Dr. Hamdi Belgacem the Chief of the Department of Electronic for his continued support during this work.

## Author

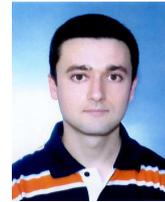

**Ahmed Rhif** was born in Sousah, Tunisia, in August 1983. He received his Engineering diploma and Master degree, respectively, in Electrical Engineering in 2007 and in Automatic and Signal Processing in 2009 from the National School of Engineer of Tunis, Tunisia (L'Ecole Nationale d'Ingénieurs de Tunis E.N.I.T). He has worked as a Technical Responsible and as a Project Manager in both LEONI and CABLITEC (Engineering automobile companies). Then he has worked as a research assistant at the Private University of Sousah (Université Privée de Sousse U.P.S) and now in the High Institute of Applied Sciences and Technologies of Sousah (Institut Supérieur des Sciences Appliquées et de Technologie de Sousse I.S.S.A.T.so). He is currently pursuing his PhD degree in the Polytechnic School of Tunis (E.P.T). His research interest includes control and nonlinear systems.
**E-mail :** ahmed.rhif@gmail.com